\documentclass[12pt]{article}

\usepackage{amssymb}
\usepackage{amsmath}

\usepackage{epsfig}

\begin{document} %

\title{\bf Randall-Sundrum model with a small curvature and dielectron production
at the LHC}

\author{A.V. Kisselev\thanks{Electronic address:
alexandre.kisselev@ihep.ru} \\
{\small Institute for High Energy Physics, 142281 Protvino, Russia}
\\
{\small and}
\\
{\small Department of Physics, Moscow State University, 119991
Moscow, Russia}}

\date{}

\maketitle

\thispagestyle{empty}


\begin{abstract}
In the framework Randall-Sundrum-like scenario with the small
curvature $\kappa$ (RSSC model), $p_{\perp}$-distributions for the
dielectron production at the LHC are calculated. For the summary
statistics taken at 7 TeV ($L = 5 \ \mathrm{fb}^{-1}$) and 8 TeV ($L
= 20 \ \mathrm{fb}^{-1}$), the exclusion limit on the 5-dimensional
gravity scale $M_5$ is found to be 6.35 TeV at 95$\%$ C.L. For
$\sqrt{s} = 13$ TeV and integrated luminosity 30 fb$^{-1}$, the LHC
search limit is estimated to be 8.95 TeV. These limits on $M_5$ are
independent of $\kappa$, provided the relation $\kappa \ll M_5$ is
satisfied.
\end{abstract}




\section{Dielectron production in the RSSC model}
\label{sec:1}

In a recent paper \cite{Kisselev:13}, the $p_{\perp}$-distributions
for the \emph{dimuon} production at the LHC were calculated in the
framework of the Randall-Sundrum-like scenario with the small
curvature (\emph{RSSC model} \cite{Giudice:05}-\cite{Kisselev:06}).
The LHC discovery limits on 5-dimensional gravity scale $M_5$ were
obtained for both 7 TeV and 14 TeV.

In contrast to the RS1 model \cite{Randall:99}, in the RSSC model
the masses of the Kaluza-Klein (KK) excitations
$h_{\mu\nu}^{(n)}(x)$ are proportional to the curvature parameter
$\kappa$ ($\kappa \ll M_5$) \cite{Kisselev:05},
\begin{equation}\label{graviton_masses}
m_n = x_n \kappa \;, \quad n=1,2, \ldots \;,
\end{equation}
where $x_n$ are zeros of the Bessel function $J_1(x)$. The
interaction of the gravitons with the SM fields is described by the
Lagrangian
\begin{equation}\label{Lagrangian}
\mathcal{L}_{\mathrm{int}} = - \frac{1}{\bar{M}_{\mathrm{Pl}}} \,
h_{\mu\nu}^{(0)}(x) \, T_{\alpha\beta}(x) \, \eta^{\mu\alpha}
\eta^{\nu\beta} - \frac{1}{\Lambda_\pi} \sum_{n=1}^{\infty}
h_{\mu\nu}^{(n)}(x) \, T_{\alpha\beta}(x) \, \eta^{\mu\alpha}
\eta^{\nu\beta} \;,
\end{equation}
where $T_{\mu \nu} (x)$ is the energy-momentum tensor of the SM
matter,
\begin{equation}\label{lambda_pi_full}
\Lambda_{\pi} \simeq \bar{M}_{\mathrm{Pl}} \, e^{-\pi \kappa r_c}
\;,
\end{equation}
and $\bar{M}_{\mathrm{Pl}}$ is the reduced Planck mass.

The goal of this paper is to estimate gravity effects in the
\emph{dielectron} production,
\begin{equation}\label{process}
p \, p \rightarrow e^+ e^- + X \;,
\end{equation}
at LHC energies in the RSSC model. The differential cross section of
the process \eqref{process} is given by
\begin{align}\label{cross_sec}
\frac{d \sigma}{d p_{\perp}}(p p \rightarrow  e^+ e^- + X) &=
2p_{\perp} \!\!\!\! \sum\limits_{a,b = q,\bar{q},g} \!\!
\int\nolimits \!\! \frac{d\tau \sqrt{\tau}}{\sqrt{\tau -
x_{\perp}^2}} \! \int\nolimits \! \frac{dx_1}{x_1}  f_{a/p}(\mu^2,
x_1) \nonumber \\
&\times f_{b/p}(\mu^2, \tau/x_1) \, \frac{d \sigma}{d\hat{t}}(a b
\rightarrow e^+ e^-) \;,
\end{align}
with the transverse energy of the electron pair equals to
$2p_{\perp}$. In eq.~\eqref{cross_sec} two dimensionless quantities
are introduced
\begin{equation}\label{tau_xtr}
x_{\perp} = \frac{2 p_{\perp}}{\sqrt{s}} \;, \quad \tau = x_1 x_2
\,,
\end{equation}
where $x_2$ is the momentum fraction of the parton $b$ in
\eqref{cross_sec}. Without cuts, integration variables in
\eqref{cross_sec} vary within the following limits
\begin{equation}\label{int_region_full}
x_{\perp}^2 \leq \tau \leq 1 \;, \quad \tau  \leq x_1 \leq 1 \;.
\end{equation}
After imposing kinematical cut on electron rapidity, the integration
region becomes more complicated (see Appendix~A in
ref.~\cite{Kisselev:13}).

The SM contribution to the $p_{\perp}$-distribution looks like
\begin{equation*}
\frac{d\sigma^{\mathrm{SM}}}{d\hat{t}}(q \bar{q} \rightarrow e^+
e^-) = \frac{1}{48 \pi s^2} \left[ u^2 \left( |G^{LL}|^2 +
|G^{RR}|^2 \right) + t^2 \left( |G^{LR}|^2 + |G^{RL}|^2 \right)
\right] \;,
\end{equation*}
with
\begin{equation*}
G^{AB}(s) = \sum_{V=\gamma,Z} \! \frac{g_A(V \rightarrow e^+e^-) \,
g_A(V \rightarrow q \bar{q})}{s - m_V^2 + i m_V \Gamma_V} \;.
\end{equation*}
Here $g_{L(R)}(\gamma \rightarrow e^+e^-) = g_{L(R)}(\gamma
\rightarrow q \bar{q}) = e$, and
\begin{align}\label{EW_couplings}
g_{L}(Z \rightarrow e^+e^-) & = -\frac{1}{2} + \sin^2 \theta_W \;, \nonumber \\
g_{R}(Z \rightarrow e^+e^-) & = \ \sin^2 \theta_W \;, \nonumber \\
g_{L}(Z \rightarrow q \bar{q}) & =T_3^q - e_q \sin^2 \theta_W \;,\nonumber \\
g_{R}(Z \rightarrow q \bar{q}) & =  - e_q \sin^2 \theta_W \;,
\end{align}
with $T_3^q$ being third component of the quark isospin, $e_q$ being
quark electric charge (in units of $|e|$).

The graviton contribution comes from both quark-antiquark
annihilation and gluon-gluon fusion subprocesses (see, for instance,
\cite{Giudice:05})
\begin{eqnarray} \label{parton_cross_sec}
\frac{d\sigma^{\mathrm{grav}}}{d\hat{t}}(q \bar{q} \rightarrow
e^+e^-) &=& \frac{\hat{s}^4 + 10\hat{s}^3 \hat{t} + 42 \, \hat{s}^2
\hat{t}^2 + 64 \hat{s} \, \hat{t}^3 + 32 \, \hat{t}^4}{1536 \, \pi
\hat{s}^2} \left| \mathcal{S}(\hat{s}) \right|^2 \;,
\nonumber \\
\frac{d\sigma^{\mathrm{grav}}}{d\hat{t}}(gg \rightarrow e^+e^-) &=&
-\frac{\hat{t}(\hat{s} + \hat{t}) (\hat{s}^2 + 2 \hat{s}\,\hat{t} +
2\,\hat{t}^2)}{256 \, \pi \hat{s}^2} \left|\mathcal{S}(\hat{s})
\right|^2
 \;,
\end{eqnarray}
where
\begin{equation}\label{KK_summation}
\mathcal{S}(s) = \frac{1}{\Lambda_{\pi}^2} \sum_{n=1}^{\infty}
\frac{1}{s - m_n^2 + i \, m_n \Gamma_n} \;
\end{equation}
is the invariant part of the partonic matrix elements, with
$\Gamma_n$ being total width of the graviton with the KK number $n$
and mass $m_n$~\cite{Kisselev:06}:
\begin{equation}\label{graviton_widths}
\Gamma_n = \eta \, m_n \left( \frac{m_n}{\Lambda_{\pi}} \right)^2,
\quad \eta \simeq 0.09 \;.
\end{equation}
Let us note that $\mathcal{S}(s)$ is a universal function for
processes mediated by $s$-channel virtual gravitons.

In the RSSC model, an explicit expression for the sum
(\ref{KK_summation}) was obtained in \cite{Kisselev:06},
\begin{equation}\label{KK_sum}
\mathcal{S}(s) = - \frac{1}{4 \bar{M}_5^3 \sqrt{s}} \; \frac{\sin 2A
+ i \sinh 2\varepsilon }{\cos^2 \! A + \sinh^2 \! \varepsilon} \;,
\end{equation}
where $\bar{M}_5 =  M_5/(2 \pi)^{1/3}$ is the reduced 5-dimensional
gravity scale, and
\begin{equation}\label{parameters}
A = \frac{\sqrt{s}}{\kappa} \;, \qquad \varepsilon = \frac{\eta}{2}
\Big( \frac{\sqrt{s}}{\bar{M}_5} \Big)^3 \;.
\end{equation}

Let us underline that the magnitude of $\mathcal{S}(s)$ is defined
by the scale $\bar{M}_5$, not by the coupling $\Lambda_{\pi}$ in the
Lagrangian \eqref{Lagrangian}. In general, this property is valid in
the RSSC model for both real and virtual production of the KK
gravitons \cite{Kisselev:05,Kisselev:06}.

\section{Numerical calculation of $\mathbf{p_{\perp}}$ distributions}
\label{sec:2}

Taking into account that the transition region $1.44 < |\eta| <
1.57$ ($1.37 < |\eta| < 1.52$) between the ECAL barrel and endcap
calorimeters is usually excluded in the CMS (ATLAS) experiment, we
impose the CMS cut on the electron pseudorapidity,
\begin{equation}\label{rapidity_cut}
|\eta| < 1.44 \;, \quad  1.57 < |\eta| < 2.50 \;.
\end{equation}
The reconstruction efficiency $85 \%$ is assumed for the dielectron
events \cite{CMS_dilepton_efficiency}. We use the MSTW NNLO parton
distributions~\cite{MSTW}, and convolute them with the partonic
cross sections. The PDF scale is taken to be equal to the invariant
mass of the electron pair, $\mu = M_{e^+e^-}$. In order to take into
account SM higher order corrections, the $K$-factor 1.5 is used for
the SM background, while a conservative value of $K=1$ is taken for
the signal.

The differential cross section of the process under consideration
has three terms
\begin{equation}\label{cross_sec_sum}
d\sigma = d \sigma (\mathrm{SM}) + d\sigma (\mathrm{grav}) + d\sigma
(\mathrm{SM}-\mathrm{grav}) \;,
\end{equation}
where the last one comes from the interference between the SM and
graviton interactions. Since the SM amplitude is pure real, while
the real part of each graviton resonance is antisymmetric with
respect to its central point, the interference term has appeared to
be negligible in comparison with the pure gravity and SM terms after
integration in partonic momenta.

The account of the \emph{graviton widths} is a crucial point for
both analytical calculations and numerical estimations. As it was
shown in refs.~\cite{Kisselev:13},\cite{Kisselev:09}, an ignorance
of the graviton widths is a \emph{rough} approximation, since it
results in very large suppression of the cross sections. The reason
lies partially in the fact that
\begin{equation}\label{gravity_cs_2}
\frac{d \sigma (\mathrm{grav})}{d p_{\perp}} \sim
\frac{1}{p_{\perp}^3} \left( \frac{\sqrt{s}}{\bar{M}_5} \right)^3
\;,
\end{equation}
while in zero width approximation one gets
\begin{equation}\label{gravity_cs_zero}
\left. \frac{d \sigma (\mathrm{grav})}{d p_{\perp}}
\right|_{\mathrm{zero \ width}} \sim \frac{1}{\bar{M}_5^3} \left(
\frac{\sqrt{s}}{\bar{M}_5} \right)^3 \;.
\end{equation}

Let us stress that in the RSSC model the gravity cross sections
\emph{do not depend} on the curvature $\kappa$ (up to small power
corrections), provided $\kappa \ll \bar{M}_5$, in contrast to the
RS1 model~\cite{Randall:99} in which all bounds on $\bar{M}_5$
depend on the ratio $\kappa/\bar{M}_{\mathrm{Pl}}$.

\begin{figure}[hbtp]
 \begin{center}
 \resizebox{7cm}{!}{\includegraphics{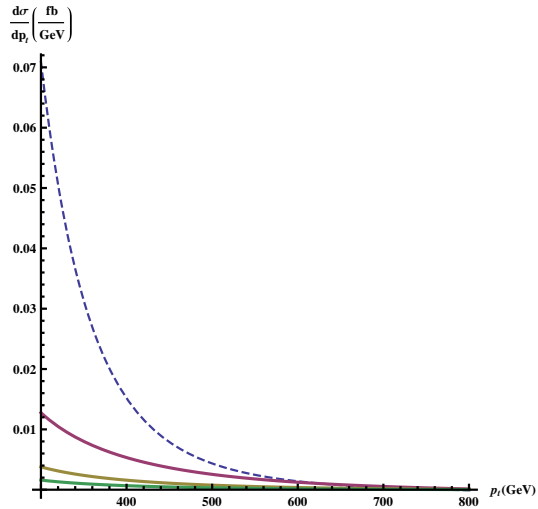}}
 \caption{The KK graviton contribution to the dielectron production for $\bar{M}_5 = 2, 4, 6$ TeV
 (solid curves, from above) vs. SM (Born) contribution (dashed curve) at $\sqrt{s} = 8$ TeV.}
 \label{fig:cs_3_8TeV}
 \end{center}
\end{figure}

\begin{figure}[hbtp]
 \begin{center}
 \resizebox{7cm}{!}{\includegraphics{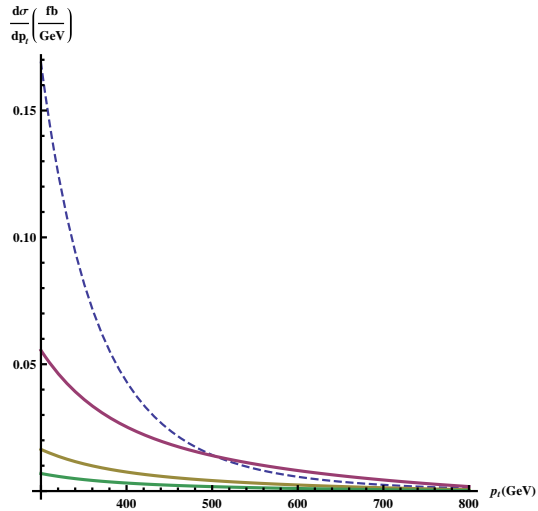}}
 \caption{The KK graviton contribution to the dielectron production for $\bar{M}_5 = 4, 6, 8$ TeV
 (solid curves, from above) vs. SM (Born) contribution (dashed curve) at $\sqrt{s} = 13$ TeV.}
 \label{fig:cs_3_13TeV}
 \end{center}
\end{figure}

In Fig.~\ref{fig:cs_3_8TeV} the results of our calculations of
gravity cross sections for the dielectron production at 8 TeV LHC
are presented.
The differential cross sections for 13 TeV are shown in
Fig.~\ref{fig:cs_3_13TeV}. Note that the gravity mediated
contributions to the cross sections do not include the SM
contribution (i.e. solid lines in the figures correspond to pure
gravity contributions).

Let $N_S$($N_B$) be a number of signal (background) dielectron
events with $p_{\perp} > p_{\perp}^{\mathrm{cut}}$,
\begin{equation}\label{ev_numder}
N_S  = \!\! \int\limits_{p_{\perp} > p_{\perp}^{\mathrm{cut}}} \!\!
\frac{d \sigma (\mathrm{grav})}{dp_{\perp}} \, dp_{\perp}  \;, \quad
N_B = \!\! \int\limits_{p_{\perp} > p_{\perp}^{\mathrm{cut}}} \!\!
\frac{d \sigma (\mathrm{SM})}{dp_{\perp}} \, dp_{\perp}
 \;.
\end{equation}

\begin{figure}[hbtp]
 \begin{center}
 \resizebox{7cm}{!}{\includegraphics{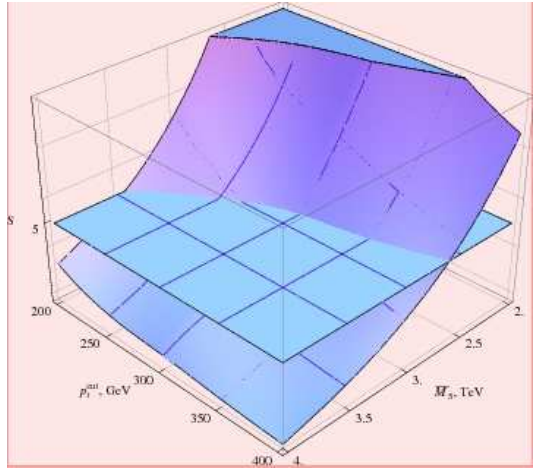}}
 \caption{The statistical significance $S$ for the dielectron
 production at the LHC for $\sqrt{s} = (7+8)$ TeV and integrated
 luminosity (5+20) fb$^{-1}$ as a function of the transverse momentum cut
 $p_{\perp}^{\mathrm{cut}}$ and \emph{reduced} 5-dimensional gravity
 scale $\bar{M}_5$. The plane $\mathcal{S}=5$ is also shown.}
 \label{fig:S_7+8TeV_5+20fb}
 \end{center}
\end{figure}

\begin{figure}[hbtp]
 \begin{center}
 \resizebox{7cm}{!}{\includegraphics{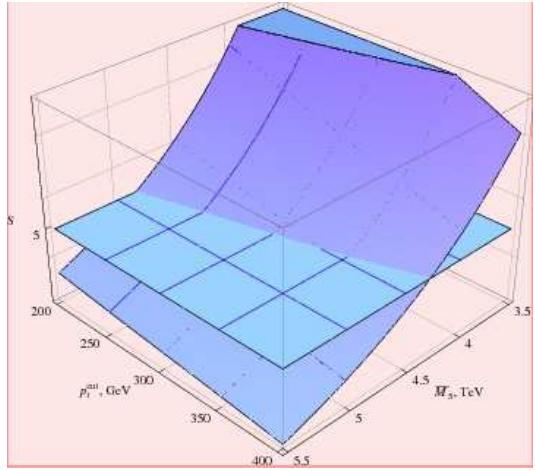}}
 \caption{The same as in figure \ref{fig:S_7+8TeV_5+20fb}, but for
 $\sqrt{s} = 13$ TeV and integrated luminosity 30 fb$^{-1}$.}
 \label{fig:S_13TeV_30fb}
 \end{center}
\end{figure}

Then we define the statistical significance $\mathcal{S} =
N_S/\sqrt{N_B + N_S}$, and require a $5 \sigma$ effect. In figure
\ref{fig:S_7+8TeV_5+20fb} the statistical significance is shown for
total number of ``events'' with $\sqrt{s} = 7$ TeV and $\sqrt{s} =
8$ TeV as a function of the transverse momentum cut
$p_{\perp}^{\mathrm{cut}}$ and \emph{reduced} 5-dimensional gravity
scale $\bar{M}_5$. The integrated luminosity was taken to be 5
fb$^{-1}$ and 20 fb$^{-1}$ for $\sqrt{s} = 7$ TeV and $\sqrt{s} = 8$
TeV, respectively.
Figure \ref{fig:S_13TeV_30fb} represents the significance
$\mathcal{S}$ for the dielectron events with $\sqrt{s} = 13$ TeV and
30 fb$^{-1}$.

Previously, calculations of dilepton cross sections were done in
\cite{Giudice:05} \emph{without} taking into account finite widths
of the KK gravitons. As was shown in \cite{Kisselev:13} (see also
\cite{Kisselev:09}), in zero width approximation the gravity cross
sections are very small in comparison with the background cross
section at low and moderate values of $p_{\perp}$. That is why, a
high cut $p_{\perp}^{\mathrm{cut}}$ is needed in order to get $N_S$
comparable with $N_B$. Correspondingly, LHC search limits in
\cite{Giudice:05} are significantly smaller than in non-zero
approximation for the graviton widths.



\section{Conclusions}
\label{sec:3}

In the present paper the RSSC model~\cite{Kisselev:05},
\cite{Kisselev:06} is considered in which the reduced 5-dimensional
Planck scale $\bar{M}_5$ is much larger that the curvature $\kappa$.
In such a model the mass spectrum and experimental signaturesare
similar to those in the ADD model~\cite{Arkani-Hamed:98} with one
flat extra dimension.

The $p_{\perp}$-distributions for the electron pairs production with
high $p_{\perp}$ at the LHC are calculated for the collision
energies 7 TeV, 8 TeV and 13 TeV, see Fig.~\ref{fig:cs_3_8TeV} and
Fig.~\ref{fig:cs_3_13TeV} (figures for 7 TeV are not shown). Let us
underline that the account of the KK graviton widths was the crucial
point for our calculations.

The statistical significance as a function of the \emph{reduced}
5-dimensional Planck scale $\bar{M}_5$ and cut on the lepton
transverse momentum $p_{\perp}^{\mathrm{cut}}$ is calculated
(Fig.~\ref{fig:S_7+8TeV_5+20fb} and Fig.~\ref{fig:S_13TeV_30fb}). No
significant deviations from the SM prediction were seen in the
dielectron events at (7+8) TeV LHC with the integrated luminosity
(5+20) fb$^{-1}$. By using our calculations, we come to the
conclusion that the region
\begin{equation}\label{search limit_7+8_TeV}
M_5 < 6.35 \mathrm{\ TeV}
\end{equation}
is excluded by experimental data. We also obtain the discovery limit
for the 13 TeV LHC with the integrated luminosity 30 fb$^{-1}$,
\begin{equation}\label{search limit_13_TeV}
M_5 = 8.95 \mathrm{\ TeV} \;.
\end{equation}

Let us stress that these bounds on $M_5$ do not depend on the
curvature $\kappa$ (up to small power-like corrections), contrary to
the RS1 model~\cite{Randall:99} in which estimated bounds on $M_5$
significantly depend on the ratio $\kappa/\bar{M}_{\mathrm{Pl}}$.



\end{document}